\documentstyle[preprint,aps,tighten]{revtex}
\begin{document}
\draft

\tolerance=5000
\def\pp{{\, \mid \hskip -1.5mm =}}
\def\cL{{\cal L}}
\def\be{\begin{equation}}
\def\ee{\end{equation}}
\def\bea{\begin{eqnarray}}
\def\eea{\end{eqnarray}}
\def\tr{{\rm tr}\, }
\def\nn{\nonumber \\}
\def\e{{\rm e}}
\def\D{{D \hskip -3mm /\,}}

\def\SEH{S_{\rm EH}}
\def\SGH{S_{\rm GH}}
\def\AdS5{{{\rm AdS}_5}}
\def\S4{{{\rm S}_4}}
\def\gfv{{g_{(5)}}}
\def\gfr{{g_{(4)}}}
\def\SC{{S_{\rm C}}}
\def\RH{{R_{\rm H}}}

\def\wlBox{\mbox{
\raisebox{0.3cm}{$\widetilde{\mbox{\raisebox{-0.3cm}\fbox{\ }}}$}}}
\def\htBox{\mbox{
\raisebox{0.1cm}{$\hat{\mbox{\raisebox{-0.1cm}{$\Box$}}}$}}}


\title{Supersymmetric new brane-world}

\author{Shin'ichi Nojiri\thanks{Electronic address: 
snojiri@yukawa.kyoto-u.ac.jp, nojiri@cc.nda.ac.jp}}
\address{Department of Applied Physics,
National Defence Academy,
Hashirimizu Yokosuka 239-8686, JAPAN}

\author{Sergei D. Odintsov\thanks{On leave from Tomsk State 
Pedagogical University, 
634041 Tomsk, RUSSIA.
Electronic address: odintsov@ifug5.ugto.mx}}
\address{
Instituto de Fisica de la Universidad de Guanajuato,
Lomas del Bosque 103, Apdo.\ Postal E-143, 
37150 Leon, Gto., MEXICO }

\maketitle
\begin{abstract}  

The quantum-induced dilatonic brane world (New Brane World) is 
created by brane CFT quantum effects (giving effective brane 
tension) in accordance with AdS/CFT set-up which also defines 
surface term. Considering the bosonic sector of 5d gauged 
supergravity with single scalar and taking the boundary action 
as predicted by supersymmetry, the possibility 
to supersymmetrize dilatonic New Brane World is discussed.
It is demonstrated that for a number of superpotentials the flat 
SUSY dilatonic brane-world (with dynamically induced brane dilaton) 
or quantum-induced de Sitter dilatonic brane-world (not Anti-de 
Sitter one) where SUSY is broken by the quantum effects occurs. 
The analysis of graviton perturbations indicates that gravity is 
localized on such branes.

\end{abstract}


\section{Introduction}

Recent increasing interest in the study of brane-worlds is caused by
several reasons. First of all, it has been realized that 4d brane 
gravity may be trapped \cite{RS1,RS2}. Second, a natural proposal 
to resolve the mass hierarchy problem appears \cite{RS1}.

An interesting variant of brane-world scenario has been suggested 
in refs.\cite{NOZ,HHR}. (It has been called New Brane 
World \cite{HHR}). In this scenario, brane-worlds are naturally 
realized in frames of AdS/CFT duality \cite{AdS}. Unlike to the 
convenient brane-worlds, the boundary action is not an arbitrary 
one (the brane tension is not a free parameter). On the contrary, 
the surface terms on AdS-like space are motivated by AdS/CFT 
correspondence. Their role is in making of variational procedure 
to be well-defined and in the elimination of the leading 
divergence of the AdS-like action. In other words, the brane 
tension is not a free parameter but it is fixed by the condition 
of finiteness of spacetime when the brane goes to infinity. 
However, in accordance with AdS/CFT correspondence there is 
quantum CFT living on the brane. Such brane quantum CFT produces 
conformal anomaly (or anomaly induced effective action) which 
leads to the creation of effective brane tension. As a result 
the dynamical mechanism to get flat or curved (de Sitter or 
Anti-de Sitter) brane-world appears \cite{NOZ,HHR} (for study 
of related questions in this scenario see \cite{ANO}) in frames 
of AdS/CFT duality. Hence, one gets less fine-tuning in 
the realization of brane-worlds as the brane tension is not 
free parameter. The nice feature of this dynamical scenario 
is that the sign of conformal anomaly terms for usual matter 
predicts de Sitter (inflationary) Universe as a preferrable 
solution in one-brane case.

The scenario of refs.\cite{NOZ,HHR} may be extended to the 
presence of non-trivial dilaton as it was done in refs.\cite{NOO}. 
Then, whole scenario looks even more related with AdS/CFT 
correspondence as dilatonic gravity naturally follows as 
bosonic sector of d5 gauged supergravity (with special 
parametrization of scalars space). The number of quantum-induced 
curved (de Sitter or Anti-de Sitter) dilatonic brane-worlds 
has been constructed in refs.\cite{NOO}.

 From another side, there is much activity now in 
supersymmetrization of Randall-Sundrum brane 
world \cite{susy1,susy2,BD,CLP} (see also refs. therein).
5d gauged supergravity represents very interesting model where 
supersymmetric dilatonic brane-world should be searched. 
Moreover, in such a model it is natural to try to construct 
supersymmetric dilatonic brane-world consistent with AdS/CFT 
correspondence \cite{AdS}. It could be then that such a scenario 
should be realized as a supersymmetric version of New Brane 
World \cite{NOZ,HHR,NOO}. 
In the present work we make an attempt in 
the construction of supersymmetric New Brane World.

In the next section the review of the construction of classical 
supersymmetric brane-world is done for bosonic sector of 5d gauged 
SG with single dilaton. Boundary action is predicted by 
supersymmetry. Half of supersymmetries survives for flat 
brane-world (as it follows from the analysis of BPS condition). 
Classical SUSY curved brane-worlds cannot be realized.
Third section is devoted to the extension of the analysis of 
second section modified by the quantum contribution from 
brane CFT in order to construct SUSY New Brane-World. It is 
shown for a number of superpotentials that unlike to the 
classical case the quantum induced de Sitter brane-world is 
created. However, the brane supersymmetry is broken by the 
quantum effects. An example of SUSY flat brane-world where 
boundary value of dilaton is defined by quantum effects is 
also given. In section four the analysis of graviton 
perturbations around found solutions is done. It is shown that 
only one normalizable solution corresponding to zero mode 
exists. In other words, gravity should be trapped on the brane 
in such scenario. Some brief summary and outlook are given in 
final section.

\section{Classical supersymmetric brane-world} 

The 5d ${\cal N}=8$ gauged supergravity can be obtained from 10d IIB 
supergravity, where the spacetime is compactified into 
S$_5\times$M$_5$. Here S$_5$ is 5d sphere and M$_5$ 
is a 5d manifold, where gauged supergravity lives. 
The bosonic sector (gravity and scalar part) of the 5d 
gauged supergravity is given by
\be
\label{Sbulk}
S_{\rm bulk}={1 \over 16\pi G}\int d^5 x \sqrt{\gfv}\left(R_{(5)} 
 -{1 \over 2}g_{ij}(\phi_k)\nabla_\mu\phi^i\nabla^\mu \phi^j 
 - V(\phi^i) \right) \ .
\ee
In the 5-dimensional maximal supergravity, the scalar field 
parametrizes the coset of $E_6/SL(6,R)$. In (\ref{Sbulk}), 
$g_{ij}(\phi^k)$ is the induced scalars metric and the 
potential $V(\phi^i)$ can be rewritten in terms of the 
superpotential $W(\phi_i)$:\footnote{Eq.(\ref{SuW0}) 
can be regarded as the definition of $W(\phi_i)$ for  
rather general potential $V(\phi)$ even if there is no 
supersymmetry. The corresponding potential in case of
 $D=5$ ${\cal N}=8$ supergravity, 
including higher rank tensors, was found in \cite{GRW}.
The potential under discussion may be considered as the one 
corresponding to some its subsector, for example, as the one discussed 
in last of refs.\cite{CLP}. 
}
\be
\label{SuW0}
V(\phi_i)={3 \over 4}\left({3 \over 2}
g^{ij}(\phi_k){dW(\phi_k) \over d\phi^i}{dW(\phi_k) \over d\phi^j}
 - W(\phi_k)^2\right)\ .
\ee
For simplicity, we may consider ${\cal N}=2$ case.
When there are boundaries or branes in 5d spacetime, 
it has been shown, for the vector multiplet, in \cite{BD} 
that the supersymmetry of the whole system consisting 
of the bulk and brane(s) is preserved by introducing 
a scalar field $\tilde G$ and four form field $A_{\mu\nu\rho\sigma}$ 
and adding the following action
\be
\label{SA}
S_A={1 \over 4!4\pi G} \int d^5 x 
\epsilon^{\mu\nu\rho\sigma\tau}A_{\mu\nu\rho\sigma}
\partial_\tau \tilde G
\ee
in the bulk spacetime and the boundary action
\be
\label{Sbndry}
S_{\rm bndry}=\mp {1 \over 8\pi G}\int d^4 x 
\left({3 \over 2}W(\phi)\sqrt{\gfr}
+ {2g \over 4!}\epsilon^{\mu\nu\rho\sigma}A_{\mu\nu\rho\sigma}
\right)
\ee
to $S_{\rm bulk}$ (\ref{Sbulk}). Here $g$ is a gauge coupling 
constant. The sign $\mp$ comes from the ambiguity when we 
solve (\ref{SuW0}) with respect to $W(\phi)$ but if there 
are two branes at $z=0$ and $z=R_b>0$, the relative sign 
should be different. 
On shell, where $\tilde G=g{\rm sgn}(z)$\footnote{
Here the function ${\rm sgn}(x)$ is defined by
\[
{\rm sgn}(x)=\left\{
\begin{array}{ll}
1 \ & x>0 \\
-1 & x<0 \\
\end{array}
\right.
\]
and $z$ is the coordinate perpendicular to the boundary or 
brane.}, the boundary action $S_{\rm bndry}$ (\ref{Sbndry}) 
has the following form:
\be
\label{Sbndry2}
S_{\rm bndry}=\mp {3 \over 16\pi G} \int d^4 x 
\sqrt{\gfr}W(\phi)\ .
\ee
This tells that the brane is BPS saturated state, that is, the 
brane preserves the half of the supersymmetries of the whole 
system. Therefore if 5d supergraviry is ${\cal N}=2$, 
4d ${\cal N}=1$ super-Yang-Mills coupled with supergravity 
would be realized on the branes. 
In order to see it, one considers the simple case that only one 
scalar field $\phi$ is non-trivial and $g_{\phi\phi}=1$ 
and investigate the equations of motion given by the simplified 
action: 
\bea
\label{S}
S&=& {1 \over 16\pi G}\left[ \int_M d^5 x
 \sqrt{-\gfv}\left( R_{(5)}
 - {1 \over 2}\partial_\mu\phi \partial^\mu\phi
 - V(\phi) \right) \right. \nn
&& \left. - \sum_{i=1,2}\int_{B_i} d^4 x \sqrt{-\gfr} 
U_i(\phi)\right]\ .
\eea
Here $B_i$'s express the boundaries or branes. 
At first, we do not specify the form of $U_i(\phi)$ but by 
investigating the equations of motion, we will see the 
correspondence with (\ref{Sbndry2}). 
We now assume the metric in 5d spacetime as 
\be
\label{Mi}
ds^2=dz^2 + \e^{2A(z)}\eta_{ij}dx^i dx^j\ ,
\ee
and $\phi$ only depends on $z$. One also supposes the branes 
sit on $z=z_1$ and $z=z_2$, respectively.
Then the equations of motion are given by
\bea
\label{Ei}
&& \phi''+ 4A'\phi' = {d V \over d \phi} + \sum_{i=1,2}
{d U_i(\phi) \over d \phi} \delta(z-z_i)\ , \\
\label{Eii}
&& 4A''+ 4(A')^2 + {1 \over 2}(\phi')^2 \nn
&& \quad = -{V \over 3}  - {2 \over 3}\sum_{i=1,2}
U_i(\phi)\delta(z-z_i) \ , \\
\label{Eiii}
&& A'' + 4 (A')^2 = {V \over 3}
 - {1 \over 6}\sum_{i=1,2}U_i(\phi)\delta(z-z_i) \ .
\eea
Here $'\equiv {d \over dz}$.
For purely bulk sector ($z_1<z<z_2$, as we assume $z_1<z_2$), 
Eqs. (\ref{Ei}-\ref{Eiii}) have the following first integrals:
\bea
\label{Iii}
\phi'={3 \over 2}{d W \over d\phi}\ ,
\quad A' = - {1 \over 4}W\ .
\eea
(Here the ambiguity in the sign when solving Eq.(\ref{SuW0}) 
with respect to $W(\phi)$ is absorbed into the definition 
of $W(\phi)$.)  One should note that classical solutions do not 
always satisfy the above Eqs. in (\ref{Iii}). Classical 
solutions are generally not invariant under the supersymmetry 
transformations and the supersymmetry in the bulk is broken 
in the classical background. Eq.(\ref{Iii}) is nothing but the 
condition that the classical solution is invariant under the 
half of the supersymmetry transformations. When there are 
branes, any solution of the equations of motion including 
the equation coming from the branes might not satisfy 
Eqs.(\ref{Iii}). We now investigate the condition for the 
brane action which allows solution satisfying Eqs.(\ref{Iii}).
Then some of the supersymmetries are preserved in the whole system. 

Near the branes, Eqs.(\ref{Ei}-\ref{Eiii}) have the following form :
\be
\label{Eiv}
\phi'' \sim {d U_i(\phi)\over d\phi}\delta (z-z_i)\ ,
\quad A'' \sim -{U_i(\phi) \over 6}\delta (z-z_i)\ ,
\ee
or
\be
\label{Eivb}
2\phi' \sim {d U_i(\phi)\over d\phi}{\rm sgn}(z-z_i)\ ,
\quad 2A' \sim -{U_i(\phi) \over 6}{\rm sgn}(z-z_i) \ ,
\ee
at $z=z_i$. If Eqs.(\ref{Iii}) are satisfied, one finds
\be
\label{Ev}
U_1(\phi)= 3W(\phi)\ ,\quad
U_2(\phi)=- 3W(\phi)\ .
\ee
Eq.(\ref{Ev}) reproduces Eq.(\ref{Sbndry2}). 
Note that Eq.(\ref{Iii}) is nothing but the BPS condition, 
where the half of the supersymmetries in the whole system 
are preserved. As we are considering the solution where 
fermionic fields vanish, the variations of the fermionic fields 
under the supersymmetry transformation should vanish if the solution 
preserves the supersymmetry. If Eq.(\ref{Iii}) is satisfied, 
the variations of gravitino and dilatino vanish under the half 
of the supersymmetry transformation.  

As an extension, one can consider the case that the brane is
curved. Instead of (\ref{Mi}), we take the following metric:
\be
\label{Mi2}
ds^2=dz^2 + \e^{2A(z)}\tilde g_{ij}dx^i dx^j\ ,
\ee
Here $\tilde g_{ij}$ is the metric of the Einstein manifold,
which is defined by
\be
\label{Ein}
\tilde R_{ij}=k \tilde g_{ij}\ ,
\ee
where $\tilde R_{ij}$ is the Ricci tensor given by
$\tilde g_{ij}$ and $k$ is a constant.
Then Eqs.(\ref{Ei}) and (\ref{Eii}) do no change but
one obtains the following equation instead of (\ref{Eiii}):
\be
\label{Eiiib}
A'' + 4 (A')^2 = k\e^{2A} + {V \over 3} 
 - {1 \over 6}\sum_{i=1,2}
U_i(\phi)\delta(z-z_i) \ .
\ee
Especialy when $k=0$,  one gets the previous solution for $\phi$,
$A$ and $U_i$. 
Even for $k=0$, the brane is not always flat, for example, if 
as $\tilde g_{ij}$ in (\ref{Mi2}), we choose the metric of the 
Schwarzschild black hole or Kerr black hole spacetime, then 
Eq.(\ref{Ein}) is satisfied since the Ricci tensor vanishes. 
 
Therefore the brane solutions with these black holes of $k=0$ 
would preserve the supersymmetry of the whole system. 
When $k\neq 0$, however, one finds that Eq.(\ref{Eiiib}) has no 
solution which satisfies the BPS condition (\ref{Iii}). 
This might tell that classical curved brane breaks the 
supersymmetry in such formalism. 
When $k>0$, the brane is 4d de Sitter space or 4d sphere when 
Wick-rotated to the Euclidean signature. On the other hand, when 
$k<0$, the brane is 4d anti-de Sitter space or 4d hyperboloid in 
the Euclidean signature.

\section{Supersymmetric new brane world}

In the previous section,  the discussion was mainly limited by flat 
brane. In this case, however, the brane crosses the event horizon 
in the finite time, which opens the causality problem. 
To avoid this problem, it would be natural to consider de Sitter 
brane which is also motivated by cosmology. If the brane is  
de Sitter space, the brane does not cross the horizon. 
Motivated by this, we consider the curved brane in this 
section, although the classical curved brane seems to break  
supersymmetry in general, as it was seen in the previous section. 

If 10d spacetime, where IIB supergraviry lives, is 
compactified into S$_5\times$M$_5$, we effectively obtain 
5d ${\cal N}=8$ gauged supergravity in the bulk and 
4d ${\cal N}=4$ $SU(N)$ or $U(N)$ super-Yang-Mills theory 
coupled with (super)gravity on the brane. On the other hand, 
if 10d spacetime is compactified into X$_5\times $M$_5$, where 
X$_5$ is S$_5/Z_2$, ${\cal N}=2$ $Sp(N)$ super-Yang-Mills 
theory coupled with (super)gravity would be realized on the 
brane. Since the matter multiplets of the super-Yang-Mills are 
coupled with (super)gravity, they generate a conformal 
anomaly on quantum level. In \cite{HHR,NOZ}, it has been 
shown that the curved brane, which is 4d de Sitter space, can be 
generated by including the conformal anomaly (via account of 
corresponding effective tension). In this way, quantum induced 
brane-worlds appear in frames of AdS/CFT duality\cite{AdS} as 
it has been explained in refs.\cite{HHR,NOZ}. Only 
non-supersymmetric configurations (for generalization 
on the non-constant dilaton presence, see \cite{NOO}) have 
been considered.
 
In this section, we include the trace anomaly induced 
action on the brane to the analysis of supersymmetric 
brane-world. One chooses the brane action to preserve 
the supersymmetry as in the previous section and consider the 
solution where the scalar field is non-trivial. 
Here  mainly Euclidean signature is used. 

As curved brane is considered, we assume that 
the metric of (Euclidean) AdS has the following form:
\be
\label{AdS}
ds^2=dz^2 + \sum_{i,j=1}^4 g_{(4)ij} dx^i dx^j\ ,
\quad g_{(4)ij}=\e^{2\tilde A(z)}\hat g_{ij}\ .
\ee
Here $\hat g_{ij}$ is the metric of the Einstein manifold as in 
(\ref{Mi2}). 
One can consider two copies of the regions given by $z<z_0$ and 
glue two regions putting a brane at $z=z_0$. 

Let us start with Euclidean signature action $S$ which is 
 sum of the Einstein-Hilbert action $\SEH$ with kinetic term 
and potential $V(\phi)$ for dilaton $\phi$, the Gibbons-Hawking 
surface term $\SGH$, the surface counterterm $S_1$ and the 
trace anomaly induced action ${\cal W}$ \cite{NOO}: 
\bea
\label{Stotal}
S&=&\SEH + \SGH + 2 S_1 + {\cal W}, \\
\label{SEHi}
\SEH&=&{1 \over 16\pi G}\int d^5 x \sqrt{\gfv}\left(R_{(5)} 
 -{1 \over 2}\nabla_\mu\phi\nabla^\mu \phi 
 - V(\phi) \right), \\
\label{GHi}
\SGH&=&{1 \over 8\pi G}\int d^4 x \sqrt{\gfr}\nabla_\mu n^\mu, \\
\label{S1}
S_1&=& -{3 \over 16\pi G l}\int d^4 x \sqrt{\gfr} W(\phi), \\
\label{W}
{\cal W}&=& b \int d^4x \sqrt{\widetilde g}\widetilde F A \nn
&& + b' \int d^4x\sqrt{\widetilde g}
\left\{A \left[2{\wlBox}^2 
+\widetilde R_{\mu\nu}\widetilde\nabla_\mu\widetilde\nabla_\nu 
 - {4 \over 3}\widetilde R \wlBox^2 
+ {2 \over 3}(\widetilde\nabla^\mu \widetilde R)\widetilde\nabla_\mu
\right]A \right. \nn
&& \left. + \left(\widetilde G - {2 \over 3}\wlBox \widetilde R
\right)A \right\} \\
&& -{1 \over 12}\left\{b''+ {2 \over 3}(b + b')\right\}
\int d^4x \sqrt{\widetilde g} 
\left[ \widetilde R - 6\wlBox A 
 - 6 (\widetilde\nabla_\mu A)(\widetilde \nabla^\mu A)
\right]^2 \nn
&& + C \int d^4x \sqrt{\widetilde g}
A \phi \left[{\wlBox}^2 
+ 2\widetilde R_{\mu\nu}\widetilde\nabla_\mu\widetilde\nabla_\nu 
 - {2 \over 3}\widetilde R \wlBox^2 
+ {1 \over 3}(\widetilde\nabla^\mu \widetilde R)\widetilde\nabla_\mu
\right]\phi \ .\nonumber
\eea 
Here the quantities in the  5 dimensional bulk spacetime are 
specified by the suffices $_{(5)}$ and those in the boundary 4 
dimensional spacetime are specified by $_{(4)}$. 

In \cite{NOO}, as an action on the brane, corresponding to 
$S_1$ in (\ref{Stotal}), the action motivated by the 
counterterm method in AdS/CFT correspondence was used:
\be
\label{S1noo}
S_1^{\rm NOO}= -{1 \over 16\pi G l}\int d^4 x \sqrt{\gfr}\left(
{6 \over l} + {l \over 4}\Phi(\phi)\right)\ .
\ee
In the AdS/CFT correspondence, the divergence coming from the 
infinite volume of AdS corresponds to the UV divergence in the 
CFT side. The counterterm which cancells the leading divergence in 
the AdS side corresponds to the above action $S_1^{\rm NOO}$.
 
In the present framework, the spacetime inside the brane has  
finite volume and there might be ambiguities when choosing the 
counterterm. Here we give $S_1$ in (\ref{Stotal}) in terms of 
the superpotential $W(\phi)$ corresponding to (\ref{SuW0}), 
which is given by 
\be
\label{SuW}
V(\phi)={3 \over 4}\left({3 \over 2}\left(
{dW(\phi) \over d\phi}\right)^2 - W(\phi)^2\right)\ .
\ee
This is natural in terms of supersymmetric extension \cite{CLP} 
of the Randall-Sundrum model \cite{RS1,RS2}. This action tells 
that the brane is BPS saturated state and the half of the 
supersymmetries could be conserved \cite{BD}. The factor $2$ 
in front of $S_1$ in (\ref{Stotal}) is coming from that we 
have two bulk regions which are connected with each other 
by the brane. 

In (\ref{GHi}), $n^\mu$ is the unit vector normal to the 
boundary. In (\ref{GHi}), (\ref{S1}) and (\ref{W}), one chooses 
the 4 dimensional boundary metric as 
\be
\label{tildeg}
\gfr_{\mu\nu}=\e^{2A}\tilde g_{\mu\nu},
\ee 
We should distinguish $A$ and $\tilde g_{\mu\nu}$ with 
$\tilde A(z)$ and $\hat g_{ij}$ in (\ref{AdS}). We will 
specify $\hat g_{ij}$ later in (\ref{metric1}). We also specify the 
quantities given by $\tilde g_{\mu\nu}$ by using $\tilde{\ }$. 

In (\ref{W}), $G$ ($\tilde G$) and $F$ ($\tilde F$) are the 
Gauss-Bonnet invariant and the square of the Weyl tensor, which 
are given as
\bea
\label{GF}
G&=&R^2 -4 R_{ij}R^{ij}
+ R_{ijkl}R^{ijkl}, \nn
F&=&{1 \over 3}R^2 -2 R_{ij}R^{ij}
+ R_{ijkl}R^{ijkl} \ ,
\eea
In the effective action (\ref{W}) induced by brane quantum 
matter, in general, with $N$ scalar, $N_{1/2}$ spinor, $N_1$ vector 
fields, $N_2$  ($=0$ or $1$) gravitons and $N_{\rm HD}$ higher 
derivative conformal scalars, $b$, $b'$ and $b''$ are 
\bea
\label{bs}
b&=&{N +6N_{1/2}+12N_1 + 611 N_2 - 8N_{\rm HD} 
\over 120(4\pi)^2}\nn 
b'&=&-{N+11N_{1/2}+62N_1 + 1411 N_2 -28 N_{\rm HD} 
\over 360(4\pi)^2}\ , 
\nn 
b''&=&0\ .
\eea
Usually, $b''$ may be changed by the finite renormalization 
of local counterterm in the gravitational effective action. 
As it was the case in ref.\cite{NOO}, the term proportional 
to $\left\{b''+ {2 \over 3}(b + b')\right\}$ in (\ref{W}), and 
therefore $b''$, does not contribute to the equations of motion. 
Note that CFT matter induced effective action may be considered 
as brane dilatonic gravity.

For typical examples motivated by AdS/CFT correspondence\cite{AdS} 
one has:


\noindent
a) ${\cal N}=4$ $SU(N)$ SYM theory 
\be
\label{N4bb}
b=-b'={C \over 4}={N^2 -1 \over 4(4\pi )^2}\ ,
\ee 
b) ${\cal N}=2$ $Sp(N)$ theory 
\be
\label{N2bb}
b={12 N^2 + 18 N -2 \over 24(4\pi)^2}\ ,\quad 
b'=-{12 N^2 + 12 N -1 \over 24(4\pi)^2}\ .
\ee
We should note that $b'$ is negative in the above cases.

Let us start the consideration of field equations. 
It is often convienient that one assumes 
the metric of 5 dimensional spacetime as follows:
\be
\label{DP1}
ds^2=g_{(5)\mu\nu}dx^\mu dx^\nu =f(y)dy^2 
+ y\sum_{i,j=1}^4\hat g_{ij}(x^k)dx^i dx^j. 
\ee
Here $\hat g_{ij}$ is the metric of 4 dimensional Einstein 
manifold as in (\ref{AdS}). A coordinate corresponding to $z$ 
in (\ref{AdS}) can be obtained by 
\be
\label{c2b}
z=\int dy\sqrt{f(y)},
\ee
and solves $y$ with respect to $z$. Then the warp
factor is $\e^{2\hat A(z,k)}=y(z)$. 

 From the variation over $g_{(5)\mu\nu}$ in the Einstein-Hilbert 
action (\ref{SEHi}), we obtain the following equation in the bulk
\bea
\label{iit}
0&=&R_{(5)\mu\nu}-{1 \over 2}g_{(5)\mu\nu}R 
 + {1 \over 2}V(\phi)g_{(5)\mu\nu} \nn
&& - {1 \over 2} \left(\partial_\mu\phi\partial_\nu\phi 
 -{1 \over 2}g_{(5)\mu\nu}g_{(5)}^{\rho\sigma}\partial_\rho \phi
\partial_\sigma \phi \right)
\eea
and from that over dilaton $\phi$
\be
\label{iiit}
0=\partial_\mu\left(\sqrt{g_{(5)}}g_{(5)}^{\mu\nu}
\partial_\nu\phi\right) - {dV(\phi) \over d\phi}\ .
\ee
Assuming that $g_{(5)\mu\nu}$ is given by (\ref{DP1}) and $\phi$ 
depends only on $y$: $\phi=\phi(y)$, we find the equations of 
motion (\ref{iit}) and (\ref{iiit}) take the following forms:
\bea
\label{DP2}
0&=&{2kf \over y}
 -{3 \over 2}{1 \over y^2} - {1 \over 2}V(\phi)f 
+ {1 \over 4}\left({d\phi \over dy}\right)^2 \\
\label{viitb}
0&=& {kf \over y} + {3 \over 4fy}{df \over dy}
 - {1 \over 2}V(\phi)f
 - {1 \over 4}\left(d\phi \over dy\right)^2 \\
\label{DP3}
0&=&{d \over dy}\left({y^2 \over \sqrt{f}}{d\phi \over dy}\right)
 - {dV(\phi) \over d\phi}y^2 \sqrt{f}\ .
\eea
Eq.(\ref{DP2}) corresponds to $(\mu,\nu)=(y,y)$ in (\ref{iit}) and 
Eq.(\ref{viitb}) to $(\mu,\nu)=(i,j)$. The case of $(\mu,\nu)=(y,i)$ 
or $(i,y)$ is identically satisfied. 

On the other hand,  brane equations are  
\bea
\label{eq2b}
0&=&{48 l^4 \over 16\pi G}\left(\partial_z A 
 - {1 \over 2}W(\phi)\right)\e^{4A}
+b'\left(4\partial_\sigma^4 A - 16 \partial_\sigma^2 A\right) \nn
&& - 4(b+b')\left(\partial_\sigma^4 A + 2 \partial_\sigma^2 A 
 - 6 (\partial_\sigma A)^2\partial_\sigma^2 A \right) \nn
&& + 2C\left(\partial_\sigma^4 \phi
 - 4 \partial_\sigma^2 \phi \right), \\
\label{eq2pb}
0&=&-{l^4 \over 8\pi G}\e^{4A}\partial_z\phi
 -{3l^3\e^{4A} \over 8\pi G}{dW(\phi) \over d\phi}\nn
&& + C\left\{A\left(\partial_\sigma^4 \phi
 - 4 \partial_\sigma^2 \phi \right) 
+ \partial_\sigma^4 (A\phi)
 - 4 \partial_\sigma^2 (A\phi) \right\}\ .
\eea
In (\ref{eq2b}) and (\ref{eq2pb}), using the coordinate $z$ 
in (\ref{c2b}) and choosing $l^2\e^{2\hat A(z,k)}=y(z)$ one 
uses the form of the metric as 
\be
\label{metric1}
ds^2=dz^2 + \e^{2A(z,\sigma)}\tilde g_{\mu\nu}dx^\mu dx^\nu\ ,
\quad \tilde g_{\mu\nu}dx^\mu dx^\nu\equiv l^2\left(d \sigma^2 
+ d\Omega^2_3\right)\ .
\ee
Here $d\Omega^2_3$ corresponds to the metric of 3 dimensional 
unit sphere. Then for the unit sphere ($k=3$)
\be
\label{smetric}
A(z,\sigma)=\hat A(z,k=3) - \ln\cosh\sigma\ ,
\ee
for the flat Euclidean space ($k=0$)
\be
\label{emetric}
A(z,\sigma)=\hat A(z,k=0) + \sigma\ ,
\ee
and for the unit hyperboloid ($k=-3$)
\be
\label{hmetric}
A(z,\sigma)=\hat A(z,k=-3) - \ln\sinh\sigma\ .
\ee
We now identify $A$ and $\tilde g$ in (\ref{metric1}) with those in 
(\ref{tildeg}). Then one finds $\tilde F=\tilde G=0$, 
$\tilde R={6 \over l^2}$ etc. 
Note that the sphere in (\ref{smetric}) corresponds to de Sitter 
space and the hyperboloid in (\ref{hmetric}) to anti-de Sitter 
space when we Wick-rotate the Euclidean signature 
to the Lorentzian one.

Using (\ref{DP2}) and (\ref{DP3}), one can delete $f$ from the 
equations and can obtain an equation that contains only the 
dilaton field $\phi$ (and, of course, bulk potential):
\bea
\label{DP4}
0&=&\left\{ {5k \over 2} - {k \over 4}y^2
\left({d\phi \over dy}\right)^2 + \left({3 \over 2}y 
 + {y^3 \over 6}\left({d\phi \over dy}\right)^2 \right) 
{V(\phi) \over 2}\right\} {d\phi \over dy} \nn
&& + {y^2 \over 2}\left({2k \over y} 
 - {1 \over 2}V(\phi)\right){d^2\phi \over dy^2}
 - \left({3 \over 4} - {y^2 \over 8}
\left({d\phi \over dy}\right)^2 \right){dV(\phi) \over d\phi}\ .
\eea
Several solutions have been found in second ref.\cite{NOO} by 
assuming the dilaton and bulk potentials as:
\bea
\label{assmp1}
\phi(y)&=&p_1\ln \left(p_2 y\right) \\
\label{assmp2}
-V(\phi)&=& c_1 \exp\left(a\phi\right) 
+ c_2\exp\left(2a\phi\right)\ , 
\eea
where $a$, $p_1$, $p_2$, $c_1$, $c_2$ are some constants. 
When $p_1=\pm {1 \over \sqrt{6}}$, Eq.(\ref{DP4}) is always 
satisfied but from Eq.(\ref{DP3}) one gets that $f(y)$ 
identically vanishes. Therefore a natural restriction is 
$p_1\neq \pm {1 \over \sqrt{6}}$. 

When $k\neq 0$, a special solution is given by
\bea
\label{case1}
&& c_1={6kp_2p_1^2 \over 3 - 2p_1^2}\ ,\quad 
c_2=0\ ,\quad a=-{1 \over p_1}\ ,\quad p_1\neq \pm \sqrt{6} \nn
&& f(y)={3- 2p_1^2 \over 4ky} \ . 
\eea
One can check that above solution satisfies (\ref{viitb}). 
Here the superpotential $W(\phi)$ is given by
\be
\label{sW2}
W(\phi)=8p_1^2\sqrt{p_2 k \over (3-2p_1^2)(8p_1^2 - 3)}
\e^{-{\phi \over 2p_1}}\ .
\ee
The potential (\ref{assmp2}) with the coefficients $c_1$ and $c_2$ 
in (\ref{case1}) corresponds to special RG flow in 5d ${\cal N}=8$ 
gauged supergravity where only one scalar from 42 scalars is 
considered. If we define $q^2$ by
\be
\label{q}
q^2\equiv {4k \over 3-2p_1^2}>0\ ,
\ee
the solution when $k=0$ can be obtained by taking $k\rightarrow 0$ 
limit and keeping $q^2$ finite. In the limit, Eqs.(\ref{case1}) 
and (\ref{sW2}) have the following forms:
\bea
\label{case1b}
&& c_1={9 \over 4}q^2 p_2\ ,\quad 
c_2=0\ ,\quad a=-{1 \over p_1}\ ,\quad p_1\neq \pm \sqrt{6} \ ,
\quad f(y)={1 \over q^2 y} \\ 
\label{sW2b}
&& W(\phi)=2\sqrt{q^2 p_2} \e^{-\phi\sqrt{3 \over 2}}\ .
\eea
This solution satisfies Eq.(\ref{Iii}), which is the BPS 
condition, i.e., the solution preserves the half of the 
supersymmetries in the bulk space.  

The solutions in (\ref{case1}) and (\ref{case1b}) have 
a singularity at $y=0$. In fact  
the scalar curvature $R_{(5)}$ is given 
by
\be
\label{Curc1}
R_{(5)} 
= - {3 \over 2}{p_1^2 q^2 \over y}\ .
\ee
Here we assume $q^2$ is defined by (\ref{q}) even if $k\neq 0$. 
When $k=3$, the brane becomes de Sitter space after the 
Wick-rotation. Then $y=0$ corresponds to the horizon in the bulk 
5d space. Therefore in $k=3$ case, the singularity is not exactly 
naked. 

In the coordinate system (\ref{DP1}), brane  
Eq.(\ref{eq2pb}) has the following form:
\be
\label{eq2pc}
0=- {y_0^2 \over 8\pi G \sqrt{f(y_0)}}\partial_y\phi 
 - {3y_0^2 \over 8\pi G}{dW(\phi_0) \over d\phi}
 + 6C\phi_0\ .
\ee
Here $\phi_0$ ($\tilde\phi_0$) is the value of the dilaton $\phi$ 
on the brane. We also find Eq.(\ref{eq2b}) has the following form:
\be
\label{eq2c1}
0= {3y_0^2 \over 16\pi G}\left({1 \over 2y_0\sqrt{f(y_0)}}
 - {l \over 2}W(\phi_0)\right) + 8b'
\ee
for $k\neq 0$ case and  
\be
\label{eq2c2}
0= {3y_0^2 \over 16\pi G}\left({1 \over 2y_0\sqrt{f(y_0)}}
 - {l \over 2}W(\phi_0)\right) 
\ee
for $k=0$ case. 

When $k=0$, where $p_1^2={3 \over 2}$, Eq.(\ref{eq2c2}) is 
satisfied trivially but Eq.(\ref{eq2pc}) has the following form:
\be
\label{eq2pc2-0}
-{qy_0^{3 \over 2} \over 8\pi G}
\sqrt{3 \over 2}= 6C\phi_0 \ .
\ee
Then the value $\phi_0$ of the dilaton on the brane depends on $y_0$. 
We should note that the obtained solution for $k=0$ is really 
supersymmetric in the whole system since the corresponding bulk 
solution (\ref{case1b}) satisfies the BPS condition 
Eq.(\ref{Iii}) which tells the solution preserves the half of the 
supersymmetries in the bulk space and the brane action has been 
chosen not to break the supersymmetry on the brane. It is 
interesting that even in case of $k=0$, the quantum effect is 
included in (\ref{eq2pc2-0}) through the parameter $C$ (coefficient 
of dilatonic term in conformal anomaly). In the classical case, 
where $C=0$, the value of the scalar field $\phi_0$ is a free 
parameter. Quantum effects suggest the way for dynamical 
determination of brane dilaton.

When $k\neq 0$, by substituting the solution in (\ref{case1}) 
into (\ref{eq2pc}) and (\ref{eq2c1}), one finds
\bea
\label{eq2pc2}
{p_1y_0^{3 \over 2} \over 4\pi G}
\sqrt{k \over 3-2p_1^2}\left(1 
 - {6 \over \sqrt{8p_1^2 - 3}}\right) &=& 6C\phi_0 \\
\label{eq2c3}
{3 y_0^{3 \over 2} \over 16\pi G}
\sqrt{k \over 3-2p_1^2}\left(1 
 - {2p_1^2 \over \sqrt{8p_1^2 - 3}}\right) &=& 
-8b' \ .
\eea
Since $b'<0$, Eqs.(\ref{eq2pc2}) and (\ref{eq2c3}) have 
non-trivial solutions for $\phi_0$ and $y_0$ if 
\be
\label{cond1}
p_1>{3 \over 8}\ ,\quad {k \over 3-2p_1^2}>0
\quad \mbox{and}\quad 1 
 - {2p_1^2 \over \sqrt{8p_1^2 - 3}}>0\ ,
\ee
The last condition in (\ref{cond1}) can be rewritten as
\be
\label{cond1l}
{1 \over 2}<p_1^2<{3 \over 2}\ .
\ee
When $k=3$, where the brane is 4d sphere (de Sitter space when 
we Wick-rotate the brane metric to Lorentzian signature), we have
\be
\label{conddS}
{3 \over 2}>p_1^2 > {7 \over 8}\ .
\ee
On the other hand, if $k=-3$, where the brane is 4d hyperboloid 
(anti-de Sitter after the Wick-rotation), there is no solution 
since the second condition in (\ref{cond1}) conflicts with 
(\ref{cond1l}). Note that de Sitter brane ($k>0$) solution 
does not exist on the classical level but the solution appeared 
after inclusion of the quantum effects of brane matter 
in accordance with AdS/CFT. 

If Eq.(\ref{conddS}) is satisfied, Eqs.(\ref{eq2pc2}) and 
(\ref{eq2c3}) can be explicitly solved with respect to 
$y_0$ and $\phi_0$. This situation is very different from 
 non-supersymmetric case in \cite{NOO}, where $S_1$ was 
chosen as in (\ref{S1noo}). In second ref. from \cite{NOO}, it 
was very difficult to solve the equations corresponding to 
(\ref{eq2pc2}) and (\ref{eq2c3}), explicitly. This indicates 
that supersymmetry simplifies the situation and the approach 
we adopt is right way to construct supersymmetric new brane world. 
Moreover, quantum effects may give a natural mechanism for 
SUSY breaking.
 
If one writes $y_0=R_b^2$, $R_b$ corresponds to the radius of the 
sphere ($k=3$). Since $b'\propto N^2$ in (\ref{N4bb}) and 
(\ref{N2bb}) for large $N$, from Eq.(\ref{eq2c3}), one gets 
\be
\label{RN}
R_b\propto \left(GN^2\right)^{1 \over 3}\ .
\ee
Note that ${1 \over R_b}$ corresponds to the expanding rate of 
the de Sitter Universe after the Wick-rotation. Therefore for 
the large quantum effect (i.e., when $N$ is large), the rate 
becomes small. 

Note that supersymmetry on the de Sitter brane should be 
broken since such spacetime is not supersymmetric background. 
On the classical level there is no de Sitter brane solution 
but only flat brane solution with $k=0$. This would tell 
that supersymmetry of the whole system does not break down 
on the classical level, even if brane is introduced. The 
quantum effects induced by the trace anomaly of brane matter 
(in accordance with AdS/CFT) could break the supersymmetry 
of the system including the brane and allow the de Sitter 
brane solution. 

\section{Gravity perturbations}

It is known that brane gravity trapping occurs on curved brane 
in a different way than on flat brane. For example, in 
refs.\cite{KR,KMP} (for dilaton presence, see last ref. 
in \cite{NOO}), the AdS$_4$ branes in AdS$_5$ were discussed 
and the existence of the massive normalizable mode of graviton 
was found. In these papers, the tensions of the branes are free 
parameters but in the case treated in the present paper the 
tension is dynamically determined. As brane solutions are 
found in the previous section when the brane is flat or de Sitter 
space, it is reasonable to consider  perturbation around 
the solution. 

Let us regard the brane as an object with a tension $\tilde U(\phi)$ 
and assume the brane can be effectively described by the 
following action, as in (\ref{S}) (for simplicity, we only 
consider the brane corresponding to $i=2$, or the limit 
$z_1\rightarrow -\infty$):
\be
\label{bten1}
S_{\rm brane}=- {1 \over 16\pi G}\int d^4x \sqrt{-g_{(4)}}
\tilde U(\phi)\ .
\ee
Then using the Einstein equation as in (\ref{Eiii}), one finds 
\be
\label{bten2}
A'' + 4 (A')^2 = -{V \over 3}
 - {\tilde U(\phi) \over 6}\delta(z-z_0) \ .
\ee
Here we assume that there is a brane at $z=z_0(=z_2)$. 
Then at $z=z_0$
\be
\label{bten3}
\left. A' \right|_{z=z_0}=-{\tilde U(\phi) \over 6}\ .
\ee
Comparing (\ref{bten3}) 
with (\ref{eq2b}) etc. one gets, when $k\neq 0$ 
\be
\label{bten4}
\tilde U(\phi) = - {3 \over l}W(\phi_0) 
+ {48\pi G b' \over R_b^4}\ .
\ee
and when $k=0$
\be
\label{bten4b}
\tilde U(\phi) = - {3 \over l}W(\phi_0) \ .
\ee
Note that tension becomes $R_b$ dependent due to the quantum 
correction when $k\neq 0$, as $b'\sim {\cal O}\left(N^2\right)$ 
and $R_b^4\sim {\cal O}\left(N^{8 \over 3}\right)$ from 
(\ref{RN}), the tension depends on $N$ as 
$\tilde U(\phi) + {3 \over l}W(\phi_0) 
\sim {\cal O}\left(N^{-{2 \over 3}}\right)$. 
One can understand that the r.h.s. in (\ref{bten4b}) and the 
first term in the r.h.s. in (\ref{bten4}) are determined 
from the supersymmetry.

We now consider the perturbation by assuming the metric 
in the following form:
\be
\label{bten5}
ds^2=\e^{2\hat A(\zeta)}\left(d\zeta^2 + \left(\hat g_{\mu\nu}
+ \e^{-{3 \over 2}\hat A(\zeta)}h_{\mu\nu}\right) dx^\mu dx^\nu\right)\ .
\ee
The following gauge conditions are chosen
\be
\label{bten6}
h^\mu_{\ \mu}=0\ ,\quad \nabla^\mu h_{\mu\nu}=0\ .
\ee 
Then one obtains the following equation
\be
\label{bten7}
\left(-\partial_\zeta^2 + {9\over 4} \left(\partial_\zeta \hat A
\right)^2 + {3 \over 2}\partial_\zeta^2 \hat A \right)h_{\mu\nu}
= m^2 h_{\mu\nu}
\ee
Here $m^2$ corresponds to the mass of the graviton on the brane 
\be
\label{bten8}
\left(\htBox + {1 \over R_b^2}\right)h_{\mu\nu} = m^2 
h_{\mu\nu}\ .
\ee
for $k>0$ and 
\be
\label{bten8b}
\htBox h_{\mu \nu} = m^2 
h_{\mu\nu}
\ee
for $k=0$. Here $\htBox$ is 4 dimensional d'Alembertian 
constructed on $\hat g_{\mu\nu}$. Since
\be
\label{bten9}
\pm\e^{A}d\zeta = dz 
= \sqrt{f}dy\ ,\quad \e^A={\sqrt{y} \over l}\ ,
\ee
one finds
\be
\label{bten10}
\pm\zeta=\int dy \sqrt{f(y) \over y} \ .
\ee
If we choose $\zeta=0$ when $y=y_0$, Eq.(\ref{bten10}) 
for the solution in (\ref{case1}) or (\ref{case1b}) gives 
\be
\label{bten12}
|\zeta|=-{1 \over q}\ln y +{1 \over q}\ln y_0\ .
\ee
Here we assume $q$ is defined by (\ref{q}) even if 
$k>0$. Since only the square of $q$ is defined in (\ref{q}), we 
can choose $q$ to be positive. 

Note that brane separates two bulk regions corresponding to 
$\zeta<0$ and $\zeta>0$, respectively.
Since $y$ takes the value in $[0,y_0]$, $\zeta$ takes the 
value in $[-\infty,\infty]$. 
Since $A={1 \over 2}\ln y$, from (\ref{bten7}) one gets
\be
\label{bten13}
\left(-\partial_\zeta^2 + {9q^2 \over 4} - 3q\delta(\zeta)
\right)h_{\mu\nu} = m^2 h_{\mu\nu}
\ee
Zero mode solution with $m^2$ of (\ref{bten13}) is given by
\be
\label{bten14}
h_{\mu\nu}=h^{(0)}_{\mu\nu}\e^{-{3q \over 2}|\zeta|}\ .
\ee
Here $h^{(0)}_{\mu\nu}$ is a constant. Any other normalizable 
solution does not exist. When 
\be
\label{min}
m^2>{9 \over 4}q^2\ ,
\ee 
there are non-normalizable solutions given by
\be
\label{bten15}
h_{\mu\nu}=a_{\mu\nu}
\cos \left(|\zeta|\sqrt{m^2-{9 \over 4}q^2}\right) 
+ b_{\mu\nu} \sin  \left(|\zeta|
\sqrt{m^2-{9 \over 4}q^2}\right) \ .
\ee
The coefficients $a_{\mu\nu}$ and $b_{\mu\nu}$ are constants of 
the integration and they are determined to satisfy the boundary 
condition, which comes from the $\delta$-function potential 
in (\ref{bten13}), 
\be
\label{bten16}
\left.{\partial_\zeta h_{\mu\nu} \over h_{\mu\nu}}
\right|_{\zeta\rightarrow 0+}
=-{3 \over 2}q\ .
\ee
Note that zero mode solution (\ref{bten14}) satisfies this 
boundary condition (\ref{bten16}). By using (\ref{bten16}), we 
can determine the coefficients $a_{\mu\nu}$ and 
$b_{\mu\nu}$ for non-normalizable solutions as follows:
\be
\label{ab}
a_{\mu\nu}=-{2b_{\mu\nu} \over 3q}\sqrt{m^2-{9 \over 4}q^2}\ .
\ee
It might be interesting that there is the minimum (\ref{min}) 
in the mass of non-normalizable mode. This situation is different 
from the original Randall-Sundrum model \cite{RS2}. Although de 
Sitter brane appears when we include the quantum correction, the 
minimum itself does not depend on the parameter of the 
quantum correction $b'$ or $N$. 

Since there is only one normalizable solution corresponding 
to zero mode (\ref{bten14}) and other solutions (\ref{bten15}) are 
non-normalizable,  gravity should be localized on the brane 
and the leading long range potential between two massive sources 
on the brane should obey the Coulomb law, i.e., 
${\cal O}\left(r^{-1}\right)$. Here $r$ is the distance between 
the above two massive sources. 
Furthermore the existence of the the minimum (\ref{min}) 
in the mass of non-normalizable mode indicates that the 
correction to the Coulomb law should be small. 

\section{Discussion}

In summary, the attempt to supersymmetrize the quantum-induced 
dilatonic New Brane World motivated by AdS/CFT is done. It is 
shown that for a number of superpotentials one can construct 
flat SUSY dilatonic brane-world or de Sitter dilatonic brane-world 
where SUSY is broken by quantum effects. The crucial role in the 
creation of de Sitter brane Universe (not Anti-de Sitter one!) is 
in account of quantum effects which produce the effective brane 
tension. The analysis of graviton perturbations for such 
brane-worlds shows that gravity trapping on the brane occurs.

It would be interesting to investigate the scenario of such sort in 
situation when not only scalar-gravitational background is 
non-trivial as in the present work but also when other 
superpartners have non-trivial background. Clearly, it is not 
so easy as the quantum effective action is getting much more 
complicated in such case. From another side, in the present 
discussion, only brane quantum effects are taken into account. 
One possibility could be to take account of bulk quantum 
effects (bulk Casimir effect \cite{GPT} in simplest 
version) in the construction of supersymmetric brane-worlds.

\noindent{\bf Acknowledgements}

We are very grateful to V.I. Tkach for numerious stimulating 
discussions.
The work of S.D.O. has been supported in part by CONACyT (CP, 
ref.990356 and grant 28454E).

\end{document}